\documentclass[12pt,preprint,dvips]{aastex}

\newcommand{\Teff}{$T_{\rm eff}$}

\newcommand{\logg}{$\log g$}

\shorttitle{Theoretical limb darkening for Cepheids II.}
\shortauthors{M. Marengo et al.}

\begin{document}

%==============================================================================
% Document Preamble
%==============================================================================

% Document title
\title{Theoretical Limb Darkening for Classical Cepheids:\\
       II. Corrections for the Geometric Baade-Wesselink Method.}

% Authors
\author{Massimo Marengo, Margarita Karovska, Dimitar D. Sasselov,
Costas Papaliolios\footnotemark[1]}
\affil{Harvard-Smithsonian Center for Astrophysics, 60 Garden St.,
Cambridge, MA 02138}
\email{mmarengo@cfa.harvard.edu}

\author{John T. Armstrong}
\affil{Remote Sensing Division, Naval Research Laboratory, Code 7210,
Washington, DC 20375}

\and

\author{Tyler E. Nordgren}
\affil{Dept. of Physics, University of Redlands, 1200 East Colton
Ave., Redlands, CA 92373}

\footnotetext[1]{We are saddened to report that our colleague, Costas
Papaliolios, died 2002 June 6.} 

% Abstract
\begin{abstract}
The geometric Baade-Wesselink method is one of the most promising
techniques for obtaining a better calibration of the Cepheid
period-luminosity relation by means of interferometric measurements of
accurate diameters. In this paper we present new wavelength- and
phase-dependent limb darkening corrections based on our time-dependent
hydrodynamic models of the classical Cepheid $\zeta$ Gem. We show that
a model simulation of a Cepheid atmosphere, taking into account the
hydrodynamic effects associated with the pulsation, shows strong
departures from the limb darkening otherwise predicted by a static
model. For most of its pulsational cycle the hydrodynamic model
predicts a larger limb darkening then the equivalent static model. The
hydrodynamics affects the limb darkening mainly at UV and optical
wavelengths. Most of these effects evolve slowly as the star pulsates,
but there are phases, associated with shocks propagating into the
photosphere, in which significant changes in the limb darkening take
place on time-scales of the order of less than a day. We assess the
implication of our model LD corrections fitting the geometric
Baade-Wesselink distance of $\zeta$~Gem for the available near-IR PTI
data. We discuss the effects of our model limb
darkening on the best fit result, and analyze the requirements needed
to test the time-dependence of the limb darkening with future
interferometric measurements. 
\end{abstract}

% Keywords
\keywords{Cepheids --- stars: atmospheres --- stars: oscillations ---
techniques: interferometric --- stars: individual ($\zeta$ Gem)}

%==============================================================================
% Document main body
%==============================================================================

%------------------------------------------------------------------------------
% Introduction
%------------------------------------------------------------------------------

\section{Introduction}\label{sec-intro}

With the recent advances in interferometry, the Baade-Wesselink (BW)
method \citep{baade1926, wesselink1946} has become one of the most
promising techniques to obtain independent distance measurements
of pulsating stars. A ``geometric'' variation of this
classical method \citep{sasselov1994} can in principle be used to
derive the distances of nearby classical Cepheids from the variations,
as they pulsate, of their angular diameter and radial velocity. The
potential accuracy of this method can improve the calibration of
the Period-Luminosity relation of Classical Cepheids
\citep{leavitt1906}, which is a fundamental step in the cosmological
distance ladder. 

The geometric BW method has been recently applied on two Cepheids in
the northern hemisphere, $\zeta$~Gem \citep{lane2000} and $\eta$~Aql
\citep{lane2002}. Both stars were observed at near-IR wavelength
(H-band) with the Palomar Testbed Interferometer (PTI). The
uniform brightness disk diameter was measured at several
epochs for both stars, with enough precision to detect the changes in
the stellar radius due to the pulsation. The measured angular
diameters were then  fitted with an appropriate model derived from the
radial velocities of the two stars, obtaining the distance with an
estimated accuracy of $\sim 10$\%.

As explained in detail by \citet{lane2002}, the main source of
uncertainty in the measurements is related to (1) the conversion of
photospheric line velocities into radial motion (the so-called
projection factor, or $p$-factor),
and (2) the estimate of the limb darkening (LD). 
These two quantities can be derived by modeling the Cepheid
atmosphere, taking into account the spherical geometry of the star and
the time-dependent hydrodynamics of the pulsations. The correct way to
determine the $p$-factor for pulsating Cepheids have been described in
detail in \citet{sabbey1995} and \citet{krockenberger1997}. 
More recently, we have presented a new
method for computing accurate time- and wavelength-dependent
center-to-limb brightness distributions for classical Cepheids
(\citealt{marengo2002}, hereafter paper~I). 

The model described in paper~I provides a significant improvement
of the limb darkening coefficients with respect to the tabulated values
currently used to analyze interferometric data. These tables
\citep{parsons1971, manduca1979, kurucz1993a, claret1995} are based on
hydrostatic model atmospheres of non-pulsating yellow supergiants,
having similar \Teff{} and \logg{} as classical Cepheids. Our models
have the advantage of being specific for each simulated
Cepheid, and provide the appropriate limb darkening for any
pulsational phase at arbitrary wavelength. As shown in paper~I, the
limb darkening predicted by our models is significantly different from
the one expected for a static yellow supergiant.

In this paper we analyze how the inclusion of hydrodynamic effects in
our Cepheid models can affect the interferometric distance
determination with the geometric BW method. We first compute the limb darkening
corrections as a function of wavelength and pulsational phase
(section~\ref{sec-LD}), starting from the center-to-limb intensity
profiles presented in paper~I for the classical Cepheid
$\zeta$~Gem. Then, we derive a new estimate for the geometric BW distance
of this star, correcting the PTI data presented by \citet{lane2002}
with our phase and wavelength dependent LD corrections
(section~\ref{sec-BW}). The accuracy of the best fit values, and the
importance of various error sources, are described in detail. We then 
conclude by analyzing the extent of the corrections induced by our
models in the geometric BW distances, and the level of accuracy that will be
required by present and future interferometers to detect the
hydrodynamic effects predicted by our models.

%------------------------------------------------------------------------------
% Limb darkening corrections
%------------------------------------------------------------------------------

\section{Limb Darkening Corrections}\label{sec-LD}

The models described in paper~I consist of a series of synthetic
atmospheres computed with second-order one-dimensional hydrodynamic
calculations, performed assuming a spherical symmetry. Each model
simulates the atmosphere of a Cepheid at a certain pulsational
phase. In paper~I we have described the general procedure to obtain
such models, and have shown the specific results for the classical
Cepheid $\zeta$~Gem. 

The pulsational period of $\zeta$~Gem (of approximately 10.15~days)
was covered by a total of 49 individual models (the time-step
in our model grid is determined by the convergence criteria in the
hydrodynamic simulation). The time resolution of our model
sequence is thus $\sim 0.2$~days. For each
model in the sequence, we have computed realistic spectral intensity
distributions. This step was done by approximating the dynamic models
with a plane-parallel atmosphere. The radiative transfer problem for
the static atmosphere was then solved in LTE conditions using the
ATLAS code \citep{kurucz1970,kurucz1979,kurucz1993b} and its opacity
library. The end result of this procedure is a set of 49 spectral
intensity distributions describing the $\zeta$~Gem spectrum as it
changes while the star pulsates.

In paper~I we showed how this procedure allows us to compute accurate
limb darkening profiles for $\zeta$~Gem, which are (pulsational) phase
and wavelength dependent. We proceed here to the next step, which is
to describe how these models can be used in interferometry. 

Interferometric measurements are usually expressed in terms of
\emph{uniform intensity disk} (UD)  or \emph{limb darkened}
diameters. The UD diameters are derived by fitting the normalized
fringe visibilities (squared) $V^2$ with a uniform disk model, which
assumes that the star is a disk of uniform brightness: 

\begin{equation}\label{eq-UD}
V^2_{UD}(\rho) = \left( \frac{2 J_1(\pi \theta_{UD} \cdot \rho)}{\pi
\theta_{UD} \cdot \rho} \right)^2
\end{equation}

\noindent
where $\theta_{UD}$ is the UD angular diameter and $\rho$ the spatial
frequency of the measurement. Since a uniform disk model is derived
assuming that the source is a disk of uniform brightness,
it does not depend on the wavelength. Real stars, however, are limb
darkened and have limb-to-center brightness distributions which are
a function of wavelength, as shown in paper~I.

For any practical purpose, we can consider the physical radius of the
stellar photosphere to be a well defined quantity.
The contribution functions of
observable spectral features in the stellar spectra does peak at
different heights in the atmosphere, but the difference is very small
compared to the full diameter of the star. For this reason, we can
assume that the angular diameter of a Cepheid is the same at all
wavelengths, after being corrected for LD. This diameter, which we
call ``limb darkened diameter'', $\theta_{LD}(\phi)$, since it takes into
account the fact that the star has LD, will thus only depend on the
pulsational phase of the star, and not on $\lambda$.
Fitting fringe visibilities with a UD model, on the other
hand, will produce a different $\theta_{UD}$ according to the
wavelength, because of the inability of UD models to take into account
the spectral properties of the LD. For this reason the UD angular
diameter measured from fringe visibilities of a LD star will instead
depend on both the pulsational phase and the wavelength:
$\theta_{UD}(\lambda,\phi)$. 

To obtain the real angular diameter of a star from the fringe
visibilities, one can either fit the $V^2$ data with an appropriate
limb darkening model, or correct the UD measurement with a specific
wavelength dependent limb darkening correction. The first approach is
certainly preferable, because preserves the wavelength dependence of
the measured visibilities, and thus allow a test the limb darkening
models. There are cases, however, in which the original visibilities
are not available, as for example when mixing heterogeneous data from
different interferometers. 

In this section we present accurate LD corrections derived using our
models. They are based on hydrodynamic models and therefore they are
phase and wavelength dependent. Even though the models 
are specific for $\zeta$~Gem, we use them here as an example to
describe the more general case of LD corrections for the generic
classical Cepheids, and their implications for interferometry and the
BW method.

The conversion factor between UD and limb darkened diameters is usually defined
as $k(\lambda,\phi) = \theta_{UD}(\lambda,\phi) /
\theta_{LD}(\phi)$. This factor is in many cases approximated by the
model of a star having similar spectral type as the source of interest
(in the case of Cepheids, a yellow supergiant, as described in
\citealt{claret1995}). Such approximations do not take into account
the phase dependence for variable sources, and may be incorrect
in their dependence on $\lambda$, since the atmosphere of a Cepheid
cannot be properly approximated with a static supergiant (see
discussion in paper~I). 

The $k(\lambda,\phi)$ corrections discussed in the following sections
are derived with the following procedure. At each phase in the model
grid, and for a set of wavelengths in the optical and near-IR range,
the center-to-limb profiles computed in paper~I are converted into
fringe visibilities by applying the Hankel transform (see
e.g. \citealt{koechlin1988}): 

\begin{equation}\label{eq-v2}
V^2_{LD}(\rho,\phi) = \left[ 2 \int^\infty_0 I_\nu(w,\lambda,\phi) \, J_0(w
\rho) \, w \, \textrm{d}w \right]^2
\end{equation}

\noindent
where $\rho$ is the spatial frequency of the interferometric
visibility, $w = \left( \theta_{LD} / 2 \right) \cdot \sin( \alpha
)$ is the projection of the stellar angular radius on the disk and
$I_\nu(w,\lambda,\phi)$ is the model stellar intensity spectrum at the
projected radius, computed for the given pulsational phase $\phi$.

The simulated $V^2_{LD}$ visibilities are then fitted with an UD
$V^2_{UD}$ model, in order to derive the UD diameter $\theta_{UD}$
which the interferometer would have measured, at each wavelength, for
the model source. The limb darkening correction $k(\lambda,\phi)$ is
the ratio between the best fit $\theta_{UD}$ and the value of the
angular diameter $\theta_{LD} = 1$~mas that we set to give a physical
dimension to the simulated visibilities in the Hankel transform. 
The simulated visibilities were computed over a spatial frequency
range up to the first minimum, to match the typical conditions
encountered when fitting real data with currently available
interferometers. We have however tried several other combinations of
model $\theta_{LD}$ and spatial frequency ranges (to simulate
different interferometer baselines), confirming that the resulting
$k(\lambda,\phi)$ are not very sensitive to these parameters, as long
as the simulated star is at least partially resolved.

%..............................................................................
% Spectral properties of limb darkening
%^^^^^^^^^^^^^^^^^^^^^^^^^^^^^^^^^^^^^^^^^^^^^^^^^^^^^^^^^^^^^^^^^^^^^^^^^^^^^^

\subsection{Spectral Properties of Limb Darkening}\label{ssec-klambda}

An example of the wavelength dependence of the LD correction
$k(\lambda)$ is shown in Figure~\ref{fig1}. The figure shows the LD
correction derived from our hydrodynamic simulations (thick line). The
thin line is the LD correction obtained from hydrostatic atmospheres 
(Kurucz models) having the \Teff$(\phi)$ and \logg$(\phi)$ measured as
a function of the pulsational phase by \citet{krockenberger1997}. 
Note that the \Teff{} of the dynamic and
static models, at each pulsational phase, is the same, since the
observational effective temperature was used as an input parameter to
compute the dynamic model. The \logg{} is instead different in the two
cases, as a consequence of the procedure followed to solve the
radiative transfer in the hydrodynamic case, and due to the specific
definition of the gravity terms in the hydrodynamic equations. 

The $k$ correction is strictly related to the spectral properties of
the source. The limb darkening is higher (lower $k$) at UV
and visible wavelengths, and converges toward unity (no limb darkening)
with increasing wavelengths toward the infrared. Spectral lines are
less limb darkened than the continuum: having higher optical depth,
they probe upper atmospheric layers, where the temperature
gradient is lower. Among the spectral features that are visible in our
low spectral resolution computations, is the Balmer jump in the UV,
the Ca H\&K doublet ($\lambda \simeq 395$~nm), H$\beta$ ($\lambda \simeq
486$~nm) and H$\alpha$ ($\lambda \simeq 656$~nm).

The first panel in Figure~\ref{fig1} shows the limb darkening
correction at minimum radius. At this phase, the Cepheid atmosphere is
most compressed, giving rise to a steeper temperature gradient. This is
responsible for an increase of the limb darkening at all wavelengths. The
LD correction for the hydrostatic atmosphere, on the other hand, does
not show this same effect, having a flatter limb darkening despite the
fact that \Teff{} is the same in the two models. The difference is as
much as $\Delta k \simeq 0.01$ at visible wavelengths, and less in the
near-IR. Accurate measurements of the LD with optical interferometers
should then be able to verify this effect. The
difference between the static and dynamic models is even larger at the
wavelengths of the main spectral features, which appear stronger in
the more compressed hydrodynamic atmosphere. 

In the next section we will show that in most cases the effects of the
hydrodynamics on the atmospheric structure result in a larger
LD. There are however phases in which the free expansion of the
atmosphere results in a \emph{quasi-static}
structure. In these phases (before and after maximum luminosity), the
hydrodynamics is less important and the two $k(\lambda)$ are virtually
indistinguishable.

As shown 
in the bottom panel, however, the full force of a shockwave crossing the
photosphere results in high excitation states and local
expansion. Even though at this phase $\zeta$~Gem is contracting (this
happens one day before minimum radius), the energy deposited by the
shock in the region where the visible photons are created generates a
lower temperature gradient. As a consequence, the LD of the
hydrodynamic simulation is lower than in the static case. Note,
however, that this effect is mostly appreciable at visible
wavelengths. An optical interferometer should detect at this phase a
decrease of the limb darkening of $\Delta k \simeq 0.01$. This phase
is very brief and, at least in the case of $\zeta$~Gem, the effects of
the shock are already dissipated after less than 20 hours, when
the LD is maximum again as the star approaches minimum radius. This
timescale is related to the shock propagation speed, which is
very well constrained by observations (see Figure~8 of
\citealt{sl1994}).

%..............................................................................
% Limb darkening and pulsational phase
%^^^^^^^^^^^^^^^^^^^^^^^^^^^^^^^^^^^^^^^^^^^^^^^^^^^^^^^^^^^^^^^^^^^^^^^^^^^^^^

\subsection{Limb Darkening and Pulsational Phase}\label{ssec-kphi}

By convolving the $k(\lambda,\phi)$ models with the filter passbands
used by interferometers, we can study the variations of the LD
correction as a function of the pulsational phase. For a convenient
comparison with observations, the best choice is to adopt the phase as
defined by the optical lightcurve, where the zero phase, $\phi_L = 0$,
coincide with maximum luminosity. As explained in paper~I, however,
our dynamic simulations are computed as a function of the phase
$\phi_V$ based on the pulsational velocity, in which zero phase is at
minimum radius. To convert our LD models to the lightcurve phases we
have used the phase shift $\Delta \phi = \phi_L - \phi_V$ between
maximum luminosity and minimum radius computed by \citet{bersier1994a,
bersier1994b}. 

Figure~\ref{fig2} shows the limb darkening corrections plotted as a
function of the visible lightcurve phase for two wavelengths. The
top panel shows $k(\phi)$ for the near-IR H filter passband
used by the PTI interferometer. The bottom panel shows
$k(\phi)$ for the bluest available channel ($\lambda \simeq 570$~nm)
of the Navy Prototype Interferometer (NPOI)
in the optical. The curves have been
filtered to remove numerical noise with an adaptive gaussian kernel,
and the 1~$rms$ error bands in our LD corrections associated to the
numerical uncertainty are indicated by two thin lines bracketing each
LD curve. 

The plots show that $k(\phi)$ is roughly constant for most phases, with
a slow increase as the star expands, as a consequence of the decreasing
temperature gradient in the photosphere. Coincident with the passage
of a shockwave through the photosphere (at $\phi_L \simeq 0.6$),
$k(\phi)$ shows a sudden rise in the optical, indicative of a sharp
decrease in the limb darkening. The energy deposited by the shock is
responsible for this effect. It increases the excitation in the
photospheric layers, and thus flattens the temperature gradient.
This extra energy dissipates in a short
timescale, after which the atmosphere resumes its normal
state. At $\phi_L \simeq 0.7$ $\zeta$~Gem is close to minimum radius,
where the temperature gradient is higher (due to the compression of
the atmosphere) and thus the star appears more limb darkened. 

The sharp increase in $k(\phi)$
observed in the optical, at the time of the shockwave, is not predicted
for the H band. This is because in this wavelength range the effects of the
shock are less pronounced. This might be due to the fact that the
emergent spectrum in the H-band originates deepest in the atmosphere
as a result of the broad minimum of the H$^-$ opacity at
1.6~$\mu$m, and the shocks increasingly steepen and disturb the
atmosphere as they propagate down the density gradient (up into
the atmosphere). 

It is important to understand how much of this behavior in the LD
curves is due to hydrodynamic effects, or just to the changes in
\Teff{} and \logg{} during the stellar pulsation. Solid lines in
Figure~\ref{fig3} show our hydrodynamic simulations for two
wavelengths. The dotted lines show the LD correction derived for
hydrostatic atmospheres having at each phase the \Teff{} and \logg{}
measured by \citet{krockenberger1997}, which we used as starting
points in our models. The overall value of $k(\phi)$ is similar, at
each wavelength, between the hydrodynamic and hydrostatic
atmospheres. The phase dependence is instead completely different.
Figure~\ref{fig4} reveals that static model LD
corrections closely follow the change of \Teff{} with phase. In a
hydrostatic atmosphere a higher \Teff{} is responsible for a flatter
photospheric temperature gradient, and thus for less limb darkening
(i.e. higher $k$). In a hydrodynamic atmosphere the temperature
gradient is determined predominantly by the pulsation dynamics, and
therefore the phase dependency of the LD is different.
In a hydrostatic atmosphere, gravity seems to have little or no effect
determining the changes in the limb darkening. When computing the
spectral energy distribution of our dynamic atmosphere, we constrain
the model \Teff{} to the observed values, leaving \logg{} as a free
parameter to be fitted with a grid of static atmospheres (see
paper~I). The rationale for this choice was that \Teff{} has a
dominant role in determining the spectral properties of the
atmosphere. This result confirms the validity of our choice. 

To test the assumption that \logg{} does plays a minor role, we have
also computed the LD correction for the static models having
\emph{both} \Teff{} and \logg{} \emph{identical} to the hydrodynamic
simulations. The results are very similar. Despite the significantly
different \logg{} with respect to the observed values, $k(\phi)$ still
closely follows the variations of \Teff. Only just after minimum
radius (when \Teff{} is maximum and the best fit \logg{} is minimum),
\logg{} plays some role in lowering the LD correction. This confirms
that, except when \logg{} is very small ($\sim 1$, in this case), the
effective temperature is the decisive parameter for the LD in a
hydrostatic atmosphere. 

In a hydrodynamic atmosphere, however, \Teff{} \emph{is not} the
dominant parameter. As shown in Figure~\ref{fig3}, when hydrodynamic
effects are taken into account, the relation between \Teff{} and the
LD breakes. Shocks are the source of the most dramatic effects in
$k(\phi)$, but even when they are absent, the LD corrections are
strikingly different from the ones predicted by hydrostatic
atmospheres. This is because they are not determined by \Teff, but by
the time-dependent structure of the expanding/contracting
atmosphere. Given that our hydrodynamic models generally have a
steeper temperature gradient than a similar hydrostatic atmosphere,
the resulting LD is larger for most phases. 

Figure~\ref{fig3} also shows the LD corrections used in
\citet{lane2002} for $\zeta$~Gem, which has the constant value of $k =
0.96 \pm 0.01$. This correction has been derived from tables published
by \citet{claret1995}, based on hydrostatic models. Note that the
average value of our LD correction is outside the error bars of the
tabulated value. The reason of this discrepancy is that our model
closely follows the \Teff{} and \logg{} derived for $\zeta$~Gem from
spectroscopic observations, while the \citet{claret1995} values are
computed for a generic grid of yellow supegiants matching the Cepheids
spectral type.

Note finally that, as expected, the LD correction is much
smaller at near-IR wavelengths, and larger
toward the blue. At bluer wavelengths the amplitude of
the $k$ variations with phase (especially at the time of the
shockwave) is also much larger. Therefore, to test
observationally the details of the hydrodynamic effects on limb
darkening,  interferometers operating at short wavelengths are
favored.

%------------------------------------------------------------------------------
% LD corrections and BW method
%------------------------------------------------------------------------------

\section{Effects of the LD corrections on the geometric BW
method}\label{sec-BW} 

The LD deviations introduced by the hydrodynamics are relatively small
(1\% in the optical, and less in the infrared), and are thus usually
ignored when treating interferometric measurements. In order to obtain
accurate distances with the BW geometric method, however, higher levels of
accuracy are needed. Under such stringent requirements, the
time and wavelength dependent hydrodynamic effects play an important
role, and should be taken into account.

A detailed description of the problems involved with the
application of the geometric BW method to pulsating Cepheids is given
in \citet{sasselov1994}. This method allows the determination of the distance
and the average radius of a pulsating star from both its angular
diameter and radial velocity at several phases. This is done by means
of a $\chi^2$ fit of the following function: 

\begin{equation}
\chi^2 = \sum_i \left[ \frac{\left( \Theta_0 + \frac{2 \Delta R_i}{D}
\right) - \theta_i}{\sigma_i} \right]^2
\end{equation}\label{eq-chi2}

\noindent
where $\Theta_0 = 2 R_0/D$ is the average angular radius of the star,
and $\Delta R_i$ its radial displacement at the time of the
interferometric measurement $\theta_i$. The variation of the stellar
radius $\Delta R(\phi_L)$  is derived, as a function of the
lightcurve phase $\phi_L$, by integrating the pulsational velocity
over time:

\begin{equation}
\Delta R(\phi_L) = - \int_{\phi_0}^{\phi_L} p \left[ v_r(\phi') -
\gamma \right] \, \hbox{d}\phi'
\end{equation}\label{eq-dR}

\noindent
where $v_r(\phi)$ is the radial velocity, which is corrected by the
systemic velocity $\gamma$ and the $p$-factor to yield the
true pulsational velocity. The systemic velocity can be derived by
requiring the conservation of the radius over one period (see
paper~I). Appropriate $p$-factors have been computed for
pulsating Cepheids such as $\zeta$~Gem; we use here the value of 1.43 (the
same adopted by \citealt{lane2002}), derived from the hydrodynamic
model we used to compute our LD profiles, which was published in
\citet{sabbey1995}.

The fit in equation~\ref{eq-chi2} can be solved analytically. The best
fit $R_0$ and $D$ can be derived by minimizing the $\chi^2$ relation
solving for the two unknown parameters. Note that we do not make
use of the color curve of the Cepheid, which can be applied to further
constrain the geometric BW solution. A detailed description of this
procedure will be given in a separate paper. 

The LD corrections $k(\phi)$ computed in the previous sections enter
the fit by allowing the conversion of the UD diameters
$\theta_{UD}^{(i)}$ obtained by the interferometer into \emph{true}
LD diameters of the star: $\theta_i = \theta_{UD}^{(i)} / k(\phi_i)$,
where $k(\phi_i)$ is the LD correction at the phase of the $i$-th
measurement, computed for the wavelength of the observation. 

The terms $\sigma_i$ in the fit $\chi^2$ relation are the errors
associated with each data point. They are the geometric sum of the
individual error sources (in terms of angular diameters) for each
observation: 

\begin{equation}
\sigma_i^2 = \left[ \sigma_i^{(\theta)} \right]^2 + \left[
\sigma_i^{(\Delta R / D)} \right]^2 + \left[ \sigma_i^{(k)} \right]^2
\end{equation}\label{eq-sigma}

\noindent
where $\sigma_i^{(\theta)}$ is the error of the interferometric
measurements, $\sigma_i^{(\Delta R / D)}$ the error related to the
radial displacement and $\sigma_i^{(k)}$ the error in the LD
correction. According to \citet{lane2002}, the errors in the PTI
H-band data vary between 0.01 and 0.06 mas. The errors on the radial
displacements are due to the uncertainty in the $p$-factor (amounting
to $\sim 4$\%, according to \citealt{sabbey1995}), and to the
estimated measurement errors in the radial velocity, adding a further
2\% according to \citet{bersier1994b}. The errors in our LD
corrections, barring systematic errors and based only on the numeric
uncertainty in our model, are of the order of $\pm 0.02$\% (see
Figure~\ref{fig2}), which is a negligible contribution with
respect to the other error sources, and is thus ignored.

The best fit results are shown in Table~1 (col. [2]) and
Figure~\ref{fig6}. Column (3) in Table~1 shows, for comparison,
the best fit parameters obtained with the same fixed LD correction
used by \citet{lane2002} of $k \simeq 0.96$. The difference in the two
best fit values of the geometric BW distance is 11~pc, which is less
than $\sim 3$\%. Even though this difference is less than 1/3 of the
error bars, it is significant: Figure~\ref{fig6} shows that the error
regions for our best fit, the UD fit and the fixed $k$ fit are mutually
exclusive. This is again a consequence of the high level of precision
that the available interferometric data already allow in the
determination of the average angular radius $\Theta_0$.

Note that the error regions shown in Figure~\ref{fig6} do not take
into account the uncertainty in the LD correction, which we ignored.
The uncertainty quoted for the tabulated LD correction used by
\citet{lane2002} is much larger than our numeric error $\Delta k \simeq
0.02$, and would have resulted in a much larger error region, including
all the narrow ovals in Figure~\ref{fig6}. The main point of this
paper, however, is to discuss the consequences of using model derived
LD corrections, and Figure~\ref{fig6} shows that in the case of
$\zeta$~Gem using generic tabulated LD corrections can lead to an
error of a few percent with respect to the best fit distance. This
discrepancy is larger than the error regions determined by all the
other error sources, and can thus in principle be tested
observationally by directly measuring the LD with longer baseline
interferometers. 

The difference in the results of the geometric BW fit are entirely due to the
LD correction, more precisely to its \emph{average value} $\bar k$
computed over the pulsational cycle. In the near-IR, and
especially in the H band, the phase dependency of $k(\lambda,\phi)$ is
relatively small (less than 2\%, compared with the error bars
$\ga 5-10$\% in the measured angular diameters). This means
that, with the current error bars in the interferometric data, we are
still not sensitive to the variations in the LD corrections induced by
the hydrodynamics. An accuracy of the order of 0.2\% in the $\theta_i$
is required to be sensitive to such effects in the H band. Note that a
more favorable situation is met in the visible, where the
temporal variations of $k(\phi)$ are larger. Given the limitation of
the available data, however, the main contribution of our hydrodynamic
simulations, at least for now, is not in the detailed dependence of
$k(\phi)$, but in setting the right level of the LD correction for the
wavelength of the observation. Contrary to the tabulated values of the
LD correction, which are computed for generic yellow supergiants
having the same spectral type of Cepheid stars, our corrections are
specific for the modeled stars, as they follow the \Teff{} and
\logg{} derived from spectroscopic observations for each star. This is
not a trivial matter, as the discrepancy with the tabulated value
($\Delta k \simeq 0.02$ in the H band) is the largest error source in
the determination of the geometric BW distance. 

Finally, we discuss the issue of the validity of the best fit
$\chi^2$ as guide to which LD better represent the data. Is the best
fit $\chi^2$ an indicator of the best value for the LD correction? If
this is the case, then we should conclude that $\zeta$~Gem is not limb
darkened, since the best fit $\chi^2$ computed with this same
procedure with the UD diameters would be $\chi^2 \simeq
27.5$. However, this is not the case.

To explain this, one should first remember that the time dependent
hydrodynamic effects are too small to be observed with the
available PTI data. While doing the BW fit, we could thus
consider the LD correction as a constant value $\bar k$ equal to its
average over the pulsational period. In the three fits shown in
Figure~\ref{fig6}, the value of $\bar k$ would be equal to 1 for the
UD data, $\sim 0.979$ for our hydrodynamic calculations and 0.96 for
Claret et al. 1995 (used by Lane et al. 2002). This coefficient enters
the $\chi^2$ equation by dividing the UD $\theta^{(i)}_{UD}$:

\begin{equation}
\chi^2 = \sum_i \left[ \frac{\left( \Theta_0 + \frac{2 \Delta
R_i}{D} \right) - \frac{\theta^{(i)}_{UD}}{\bar k}}{\sigma_i} \right]^2
\end{equation}\label{eq-chi2-thk}

A smaller value of $\bar k$ (larger LD), will result in a
larger best fit mean angular diameter $\Theta_0$, which will thus
scale as $1/\bar k$. The net effect of this scaling is that the best
fit $\chi^2$ will scale as $1/\bar k$. This means that a fit made with
larger LD (as long as the phase dependency of $k$ is irrelevant) will
have a larger best fit $\chi^2$. The UD fit will thus appear always to
be the best. 

The conclusion is that, until the accuracy of the measured
angular diameters becomes good enough to appreciate the changes in LD
due to the pulsation, the geometric BW fit is not a good tool to test the LD
models. On the other hand, a good model for the LD is absolutely
necessary to obtain a reliable value of the geometric BW distance,
since the best fit angular diameter scales linearly as $1/\bar k$.

Note, finally, that a different fitting strategy may allow a test our
LD models even with present day interferometric data. Our
insensitivity to the LD is due to the fact that we are fitting derived
data (the angular diameters $\theta_i$) which have been previously
obtained from the fringe visibilities \emph{assuming} a uniform disk
model. This initial fit has destroyed the sensitivity of the original
data to the stellar LD. The correct way to test the LD is to perform
$\chi^2$ fits on the original visibilities. Any test for the absolute
value of LD predicted by our models should thus be made using the
visibility data taken at different projected baselines and
wavelengths.

%------------------------------------------------------------------------------
% Conclusions
%------------------------------------------------------------------------------

\section{Conclusions}\label{sec-concl}

The geometric BW method is a powerful tool to derive the distances of
pulsating stars. The detailed analysis of the method, and our
discussion of its main current uncertainties described in the previous
sections, show that a special care should be taken when assessing
the accuracy of the results.

Despite the increasing quality of the interferometric data, the error
bars in the individual measurements are the largest contributors
to the final errors in the geometric BW distances and average radii. After 
the instrumental errors, however, follows the uncertainty in the limb
darkening correction. The importance of LD will grow in the near
future, given the fast pace at which the available interferometers are
improving their accuracy, and with the new long baseline, large
aperture interferometers which are becoming operational.
To address this issue, we have presented in this paper a
procedure to compute accurate LD corrections for pulsating Cepheids,
which are based on time- and wavelength-dependent hydrodynamic
models. 

We show that our $k(\lambda,\phi)$ strongly differ
from the equivalent corrections computed from hydrostatic atmospheres,
even in absence of ``strong'' hydrodynamic effects like shocks. The
main consequence of the hydrodynamic terms in our stellar atmospheric
models is an increase in the limb darkening,  because of a generally
higher temperature gradient in the photosphere. The situation briefly
reverses in presence of shockwaves, due to the energy deposited by the
shocks. 

The magnitude of these effects depends on the wavelength. The
effect is more significant in the
visible spectrum ($\Delta k \sim 0.015$) and it is smaller
in the infrared ($\Delta k \sim 0.002$). Current
interferometers still cannot test these time-dependent variations in
the limb darkening, which will however become important in the near
future, especially when interferometric observations in spectral lines
will become feasible. The phases of rapid variation in the LD
associated to the propagation of the shockwave in the photosphere,
when significant changes in the LD occur in timescales of hours,
appears particularly appealing for an observational test of our
models. 

Even though the time-dependent variations in the LD are not currently
measurable because of the required visible accuracy,
our models can already provide an average value of
the wavelength-dependent LD correction $k(\lambda)$ which is
significantly different from the tabulated values currently used. In
the case of $\zeta$~Gem, our limb darkening corrections induce a
$\sim 3$\% change in the best fit value of the BW distance derived
from H-band PTI data. Even more important is that, as shown in
section~\ref{sec-BW}, the $1\sigma$ error regions around the best
fit distances derived with our and other LD corrections are mutually
exclusive, opening the possibility of an independent test of our
models when longer baselines reaching the first minimum in the
visibilities will allow a direct test. A preliminary test of
our models will also be possible by directly fitting the observed
visibilities with our model visibilities, instead of using LD
corrections of UD best fit diameters.

As the data for other Cepheids will become available, it will be
important to have a large library of models specifically computed for
each source. As the models are very dependent on the pulsational
characteristics of each star, upon which the hydrodynamic model is
build, a ``generic'' parametric limb darkening correction for all
stars is not possible, as any individual model cannot be extended to
be used for a Cepheid with a different pulsational engine. This is the
reason why tables of LD corrections as the ones produced by
\citet{claret1995} cannot reproduce the detailed changes in the LD
which are required to apply the geometric BW method to classical
Cepheids.

%==============================================================================
% Document References and floating stuff
%==============================================================================

% Acknowledgements
\acknowledgements
We wish to thank the anonymous referee and the editor for the comments
and suggestions that helped us to improve this paper.
This work was partially supported by NFS grant AST 98-76734. M.K. is
a member of the Chandra Science Center, which is operated under
contract NAS8-39073, and is partially supported by NASA.

%------------------------------------------------------------------------------
% The bibliography
%------------------------------------------------------------------------------

\clearpage

%==============================================================================
% Tables
%==============================================================================

% Table 1 (LD fit)

\begin{table}
\begin{center}
\begin{tabular}{lcc}
\multicolumn{3}{c}{TABLE 1}\\
\multicolumn{3}{c}{BW BEST FIT OF $\zeta$~GEM PTI H-BAND LD DATA}\\
\hline
\hline
Fit Parameter     & $k(\phi)$ (model)      & $k = 0.96$ \\
(1)               & (2)                    & (3) \\
\hline
$R_0$ [R$_\odot$] & 67.0$^{+8.7}_{-6.9}$   & 66.2$^{+8.3}_{-6.6}$ \\
$D$ [pc]          & 372$^{+49}_{-39}$      & 361$^{+46}_{-36}$ \\
$\Theta_0$ [mas]  & 1.665$\pm 0.007$       & 1.698$\pm 0.007$ \\
$\chi^2$          & 28.1                   & 29.5 \\
\hline
\end{tabular}
\end{center}
\end{table}

\clearpage

%------------------------------------------------------------------------------
% Figure captions
%------------------------------------------------------------------------------

\figcaption[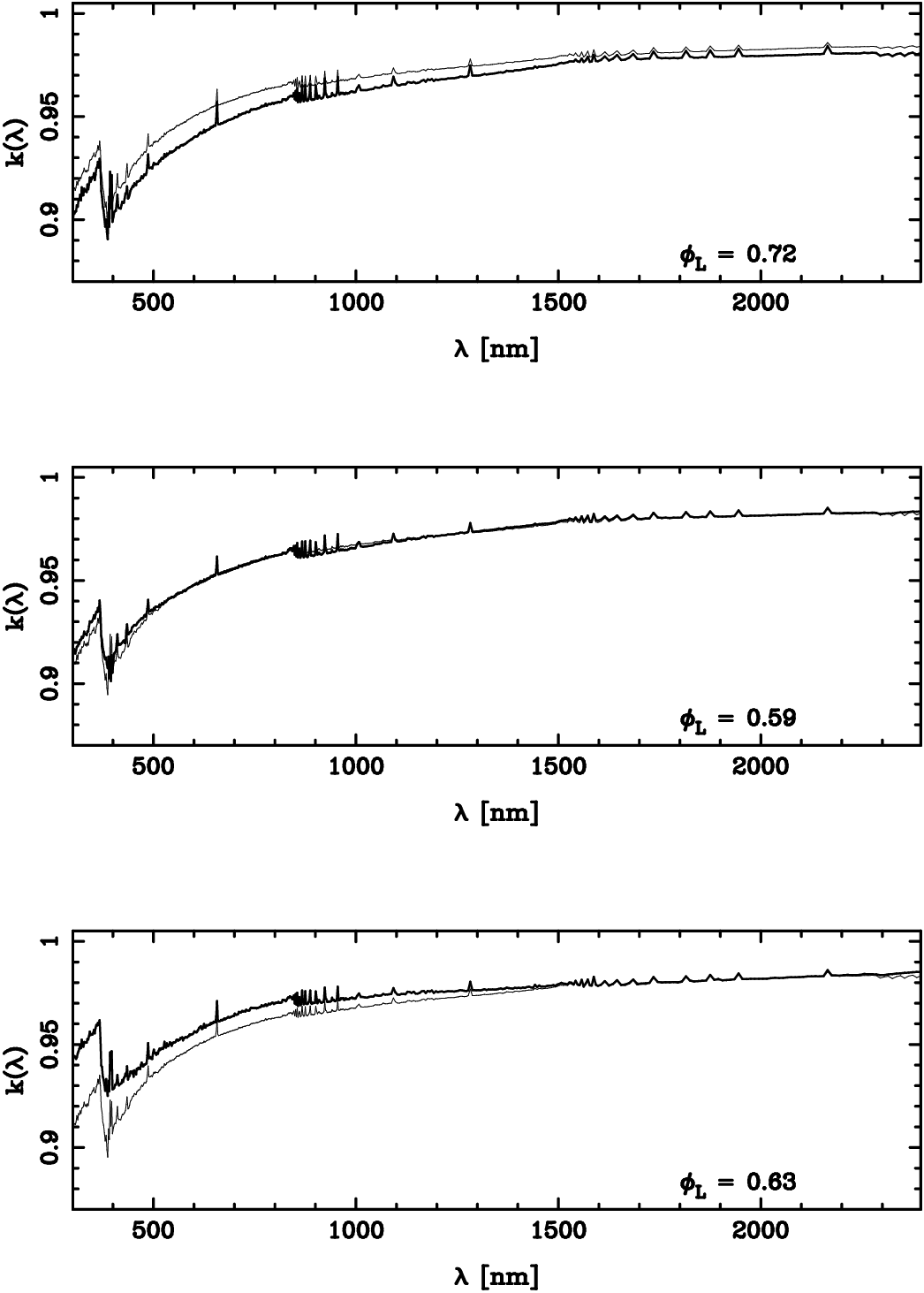]{Wavelength dependence of the LD correction
$k(\lambda)$ for our $\zeta$~Gem model. From top to bottom, the
models are shown at minimum radius, at a quasi-static phase after
maximum luminosity, and at the time in which a shockwave is crossing
the atmosphere. Solid thick line is $k(\lambda)$ from our hydrodynamic
model, while the thin lines are derived from a hydrostatic model
having the \Teff{} and \logg{} determined from observations
\citep{krockenberger1997}.\label{fig1}}

\figcaption[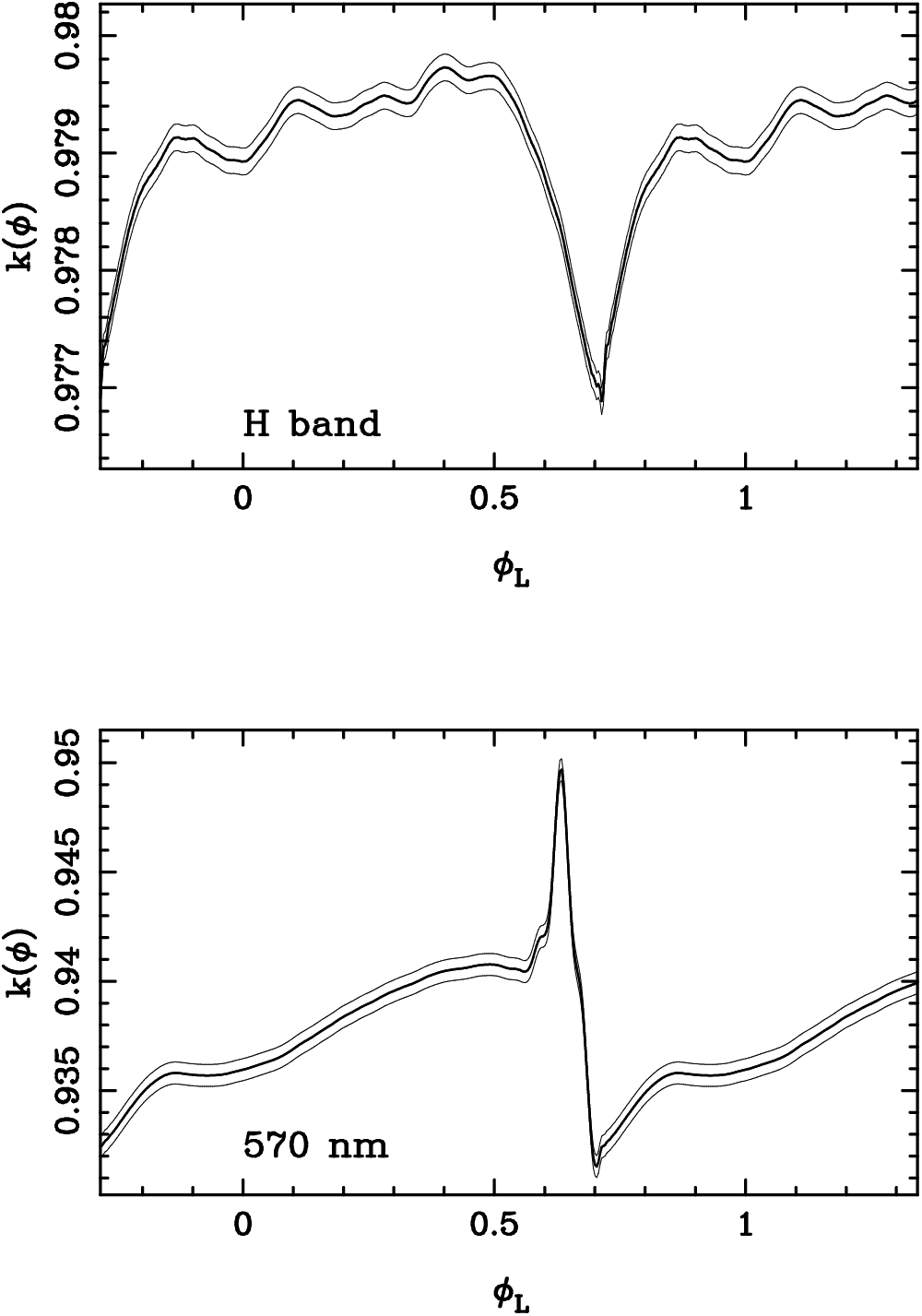]{Pulsational phase dependence of the LD correction
$k(\phi)$ for our $\zeta$~Gem model. The LD
correction is shown for the near-IR H band (top), and
at visible wavelength (570~nm, bottom). The thin lines are the 1~$rms$
numerical uncertainties in our simulations.\label{fig2}}

\figcaption[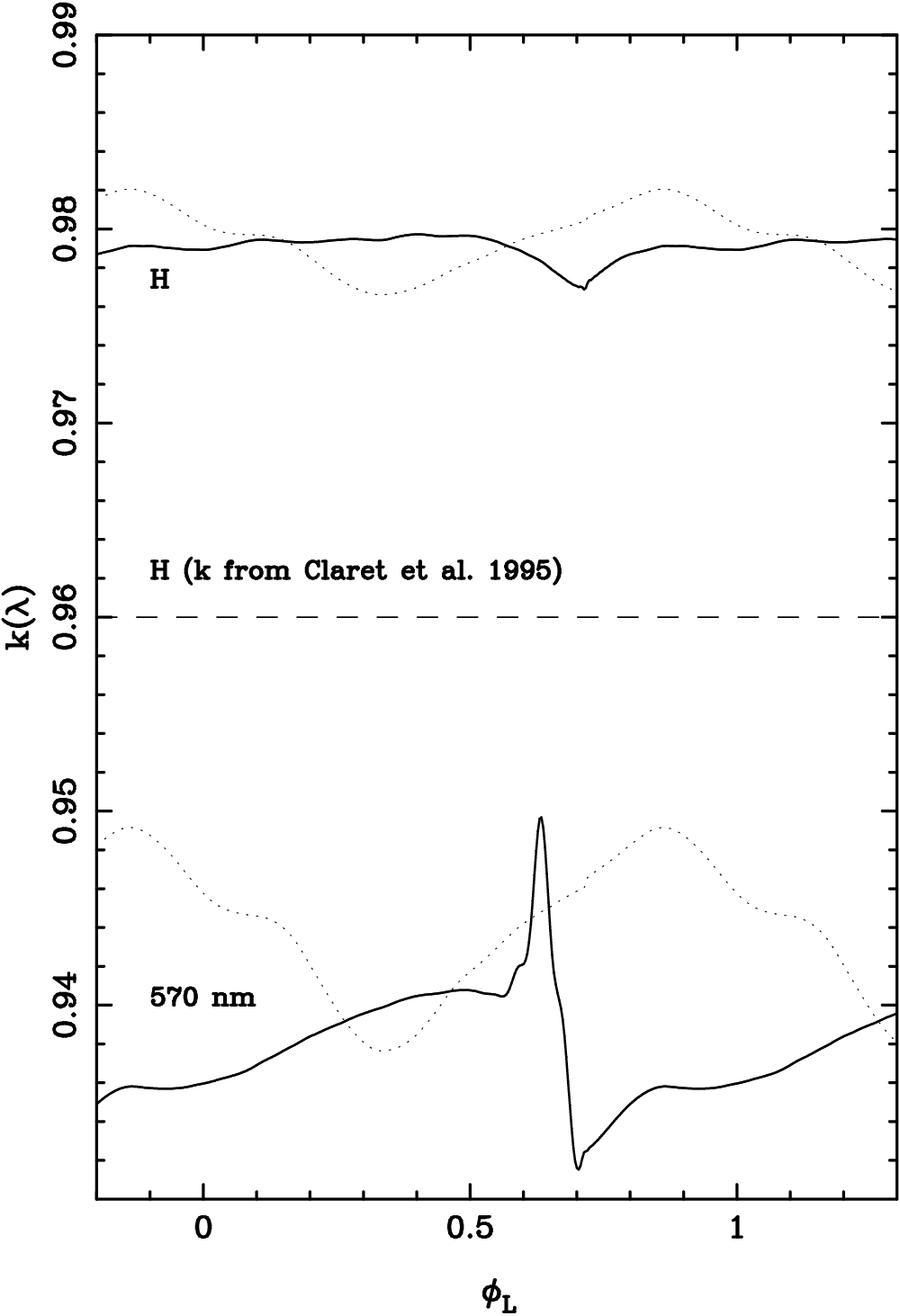]{Phase dependent $k(\phi_L)$ for $\zeta$~Gem in
the PTI H band, and in the bluest 570~nm channel of
NPOI. Solid lines are the LD corrections computed with our
hydrodynamic model; dotted lines are the equivalent corrections for an
hydrostatic atmosphere having \Teff{} and \logg{} from
\citet{krockenberger1997}. The dashed line is the value computed by
\citet{claret1995} and used in \citet{lane2002}.\label{fig3}} 

\figcaption[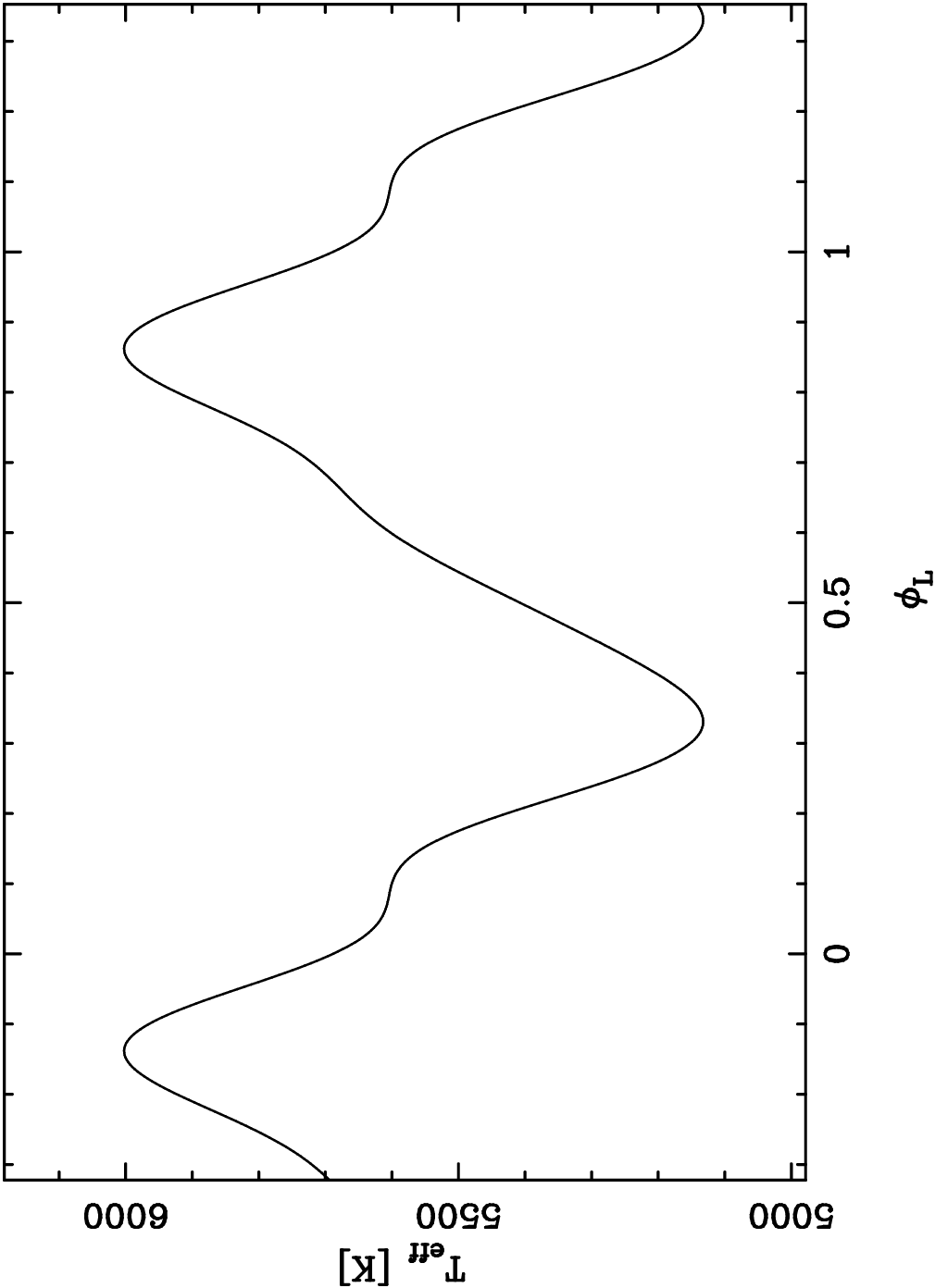]{Effective temperature, as a function of lightcurve
phase, used to compute the hydrodynamic model of $\zeta$~Gem. The
dependence of \Teff{} from the pulsational phase is derived from
measurements by \citet{krockenberger1997}.\label{fig4}} 

\figcaption[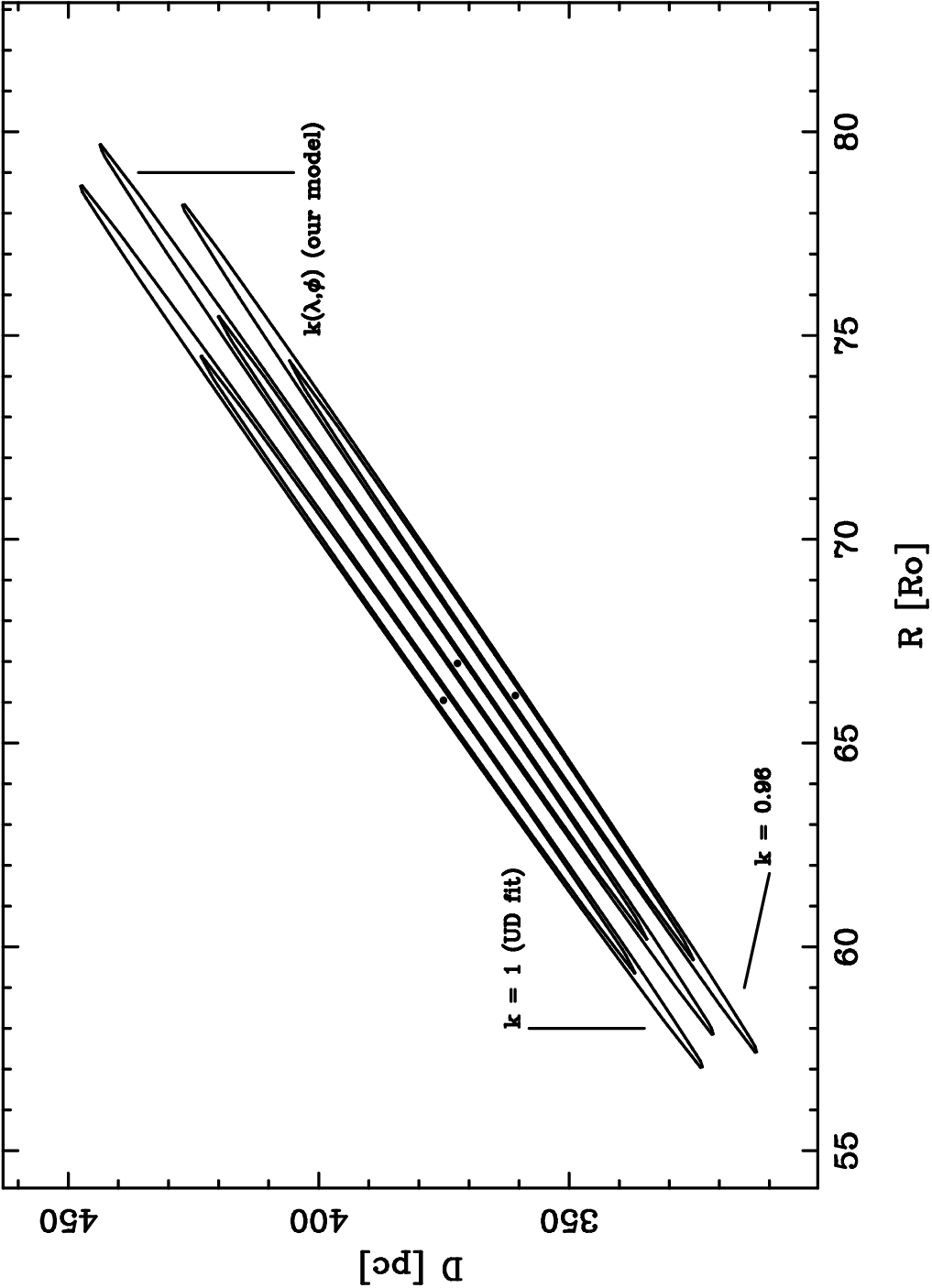]{Best fit parameters and error regions for
$\zeta$~Gem PTI H-band data. Top curve is the UD data, middle curve is
our best fit for model LD data and the bottom curve is the result
obtained using a fixed LD correction of $k \simeq 0.96$ as in
\citet{lane2002}. The inner error region is the 68\% confidence level
of the fit (1$\sigma$), while the outer is the 90\% confidence
level.\label{fig6}}

%------------------------------------------------------------------------------
% Include figures
%------------------------------------------------------------------------------

\epsscale{0.85} \plotone{f1.eps} \clearpage
\epsscale{0.85} \plotone{f2.eps} \clearpage
\epsscale{0.85} \plotone{f3.eps} \clearpage
\epsscale{0.85} \plotone{f4.eps} \clearpage
\epsscale{0.85} \plotone{f5.eps} \clearpage

% End of document
\end{document}